\documentclass[showkeys,showpacs,superscriptaddress]{revtex4-2}
\usepackage{amsmath}
\usepackage{amssymb}
\usepackage{graphicx}
\usepackage{bm}
\usepackage{float}
\usepackage{ulem,xcolor}
\usepackage{color,soul}

\begin{document}

\title{Mass gap for a monopole interacting with different nonlinear spinor fields
}

\author{
Vladimir Dzhunushaliev
}
\email{v.dzhunushaliev@gmail.com}
\affiliation{
Institute of Nuclear Physics, Almaty 050032, Kazakhstan
}

\affiliation{
Department of Theoretical and Nuclear Physics,  Al-Farabi Kazakh National University, Almaty 050040, Kazakhstan
}
\affiliation{
Academician J.~Jeenbaev Institute of Physics of the NAS of the Kyrgyz Republic, 265 a, Chui Street, Bishkek 720071, Kyrgyzstan
}

\author{Vladimir Folomeev}
\email{vfolomeev@mail.ru}
\affiliation{
Institute of Nuclear Physics, Almaty 050032, Kazakhstan
}
\affiliation{
Academician J.~Jeenbaev Institute of Physics of the NAS of the Kyrgyz Republic, 265 a, Chui Street, Bishkek 720071, Kyrgyzstan
}
\affiliation{
Laboratory for Theoretical Cosmology, International Centre of Gravity and Cosmos,
Tomsk State University of Control Systems and Radioelectronics (TUSUR),
Tomsk 634050, Russia
}

\author{
 Aizhan Abdykaliyeva
}
\affiliation{
Department of Theoretical and Nuclear Physics,  Al-Farabi Kazakh National University, Almaty 050040, Kazakhstan
}

\author{
	Symbat Khussainova
}
\affiliation{
	Department of Theoretical and Nuclear Physics,  Al-Farabi Kazakh National University, Almaty 050040, Kazakhstan
}

\author{
	Akbota Temirova
}
\affiliation{
	Department of Theoretical and Nuclear Physics,  Al-Farabi Kazakh National University, Almaty 050040, Kazakhstan
}

\begin{abstract}
We study the properties of the mass gap for monopole solutions in SU(2) Yang-Mills theory with a source of the non-Abelian gauge field
in the form of a spinor field described by the nonlinear Dirac equation. Different types of nonlinearities parameterized by the parameter $\lambda$
are under consideration. It is shown that for the range of values of this parameter studied in the present paper the value
of the mass gap depends monotonically on this parameter, and the position of the mass gap does not practically depend on $\lambda$.
\end{abstract}

\pacs{12.38.Mh, 11.15.Tk, 12.38.Lg, 11.15.-q
}

\keywords{
non-Abelian SU(2) theory; nonlinear Dirac equation; monopole;  energy spectrum; mass gap; nonlinearity parameter
}
\date{\today}

\maketitle

\section{Introduction}

The problem of a mass gap occupies a special place in quantum chromodynamics.  The significance of this problem
is related to the fact that its solution requires the development of nonperturbative quantization methods in field theory, which are absent at present.
It should be emphasised that the nonperturbative quantization is carried out numerically in lattice calculations. However, such calculations
do not enable one to have a full understanding of the nonperturbative quantization, and some questions of principle still remain to be clarified:
What are the properties of operators of a strongly interacting field? How one can determine a quantum state of such a field? Also,
there are many other unclear questions for which there are answers in the case of perturbative quantization using Feynman  diagrams.

In this connection, the question arises as to the existence of a mass gap in other theories. The presence of a mass gap
in any theory is of interest by itself, but it is also possible that such a theory may serve as some approximate description of nonperturbative effects
in quantum theory involving a strong interaction. In Refs.~\cite{Dzhunushaliev:2020qwf,Dzhunushaliev:2021apa}, it is shown that in
SU(2) gauge theory with a source in the form of a nonlinear spinor field, there exist topologically trivial solutions with a radial magnetic field
(``monopoles''). Such ``monopoles'' possess the following features:  (a)~the radial field decreases as  $r^{-3}$ at infinity
(this is the reason why we use quotation marks for the word monopole); (b)~the solutions are  topologically trivial, unlike those describing the 't~Hooft-Polyakov monopole;
(c)~the most interesting fact is that the energy spectrum of such solutions has an absolute minimum, which we can refer to as a mass gap.

In the absence of a SU(2) gauge field, the corresponding mass gap was found in Refs.~\cite{Finkelstein:1951zz,Finkelstein:1956} in studying a nonlinear spinor field.
These references also consider a generalization of  nonlinear spinor field when one introduces some constant in the nonlinear term for the spinor field.
In the present paper we study effects of the presence of a mass gap and its properties in a system containing such a nonlinear spinor field.

The nonlinear Dirac equation was introduced by W.~Heisenberg as an equation describing the properties of an electron. The classical properties  of this equation,
in particular, its soliton-like solutions, were investigated in Refs.~\cite{Finkelstein:1951zz,Finkelstein:1956}. Subsequently one of variants of the nonlinear Dirac equation
was employed for an approximate description of the properties of hadrons (this approach is called the Nambu-Jona-Lasinio model~\cite{Nambu:1961tp}; for a review, see Ref.~\cite{Klevansky:1992qe}).

In the present paper we study the dependence of the size of a mass gap, its position, etc., on the value of a parameter describing the nonlinear self-interaction potential of the spinor field.

\section{Equations and Ans\"{a}tze for Yang-Mills fields coupled to a nonlinear Dirac field}
\label{YM_Dirac_scalar}

In this section we closely follow Ref.~\cite{Dzhunushaliev:2020qwf,Dzhunushaliev:2021apa}. The Lagrangian describing a system consisting of a non-Abelian SU(2) field $A^a_\mu$
interacting with nonlinear spinor field $\psi$ can be taken in the form
\begin{equation}
\begin{split}
	\mathcal L = & - \frac{1}{4} F^a_{\mu \nu} F^{a \mu \nu}
	+ i \hbar c \bar \psi \gamma^\mu D_\mu \psi  -
	m_f c^2 \bar \psi \psi
	+ \frac{\Lambda}{2} g \hbar c V \left( \bar \psi, \psi \right).
\label{1_10}
\end{split}
\end{equation}
Here $m_f$ is the mass of the spinor field;
$
D_\mu = \partial_\mu - i \frac{g}{2} \sigma^a
A^a_\mu
$ is the gauge-covariant derivative, where $g$ is the coupling constant and $\sigma^a$ are the SU(2) generators (the Pauli matrices);
$
F^a_{\mu \nu} = \partial_\mu A^a_\nu - \partial_\nu A^a_\mu +
g \epsilon_{a b c} A^b_\mu A^c_\nu
$ is the field strength tensor for the SU(2) field, where $\epsilon_{a b c}$ (the completely antisymmetric Levi-Civita symbol) are the SU(2) structure constants;  $\Lambda$ is a constant; $\gamma^\mu$ are the Dirac matrices in the standard representation; $a,b,c=1,2,3$ are color indices and $\mu, \nu = 0, 1, 2, 3$ are spacetime indices; $V \left( \bar \psi, \psi \right)$
is the potential describing the nonlinear self-interaction of the spinor field $\psi$. According to Ref.~\cite{Finkelstein:1956}, this potential is a linear
combination of scalar and pseudoscalar interactions:
\begin{equation}
	V \left( \bar \psi, \psi \right) = \alpha \left( \bar \psi \psi \right)^2 +
	\beta \left( \bar \psi \gamma^5 \psi \right)^2,
\label{1_15}
\end{equation}
where $\alpha$ and $\beta$ are constants.


Our purpose is to study monopole-like solutions of these equations. To do this, we use the standard SU(2) monopole {\it Ansatz}
\begin{eqnarray}
	A^a_i &=&  \frac{1}{g} \left[ 1 - f(r) \right]
	\begin{pmatrix}
		0 & \phantom{-}\sin \varphi &  \sin \theta \cos \theta \cos \varphi \\
		0 & -\cos \varphi &   \sin \theta \cos \theta \sin \varphi \\
		0 & 0 & - \sin^2 \theta
	\end{pmatrix} , \quad
	i = r, \theta, \varphi  \text{ (in polar coordinates)},
	\label{2_10}\\
	A^a_t &=& 0 ,
	\label{2_13}
\end{eqnarray}
and the {\it Ansatz} for the spinor field from Refs.~\cite{Li:1982gf,Li:1985gf}
\begin{equation}
	\psi^T = \frac{e^{-i \frac{E t}{\hbar}}}{g r \sqrt{2}}
	\begin{Bmatrix}
		\begin{pmatrix}
			0 \\ - u \\
		\end{pmatrix},
		\begin{pmatrix}
			u \\ 0 \\
		\end{pmatrix},
		\begin{pmatrix}
			i v \sin \theta e^{- i \varphi} \\ - i v \cos \theta \\
		\end{pmatrix},
		\begin{pmatrix}
			- i v \cos \theta \\ - i v \sin \theta e^{i \varphi} \\
		\end{pmatrix}
	\end{Bmatrix},
	\label{2_20}
\end{equation}
where $E/\hbar$ is the spinor frequency and the functions $u$ and $v$ depend on the radial coordinate $r$ only.

We will seek a solution to the Euler-Lagrange equations coming from the variation of the Lagrangian~\eqref{1_10}. Substituting in these equations
the {\it Ans\"atz}~\eqref{2_10}-\eqref{2_20}, the expression~\eqref{1_15}, and integrating over the angles
$\theta, \varphi$ (for details see Ref.~\cite{Finkelstein:1956}), the resulting
equations for the unknown functions $f, u$, and $v$ are
\begin{eqnarray}
	- f^{\prime \prime} + \frac{f \left( f^2 - 1 \right) }{x^2} +
	\tilde g^2_{\Lambda} \frac{\tilde u \tilde v}{x} &=& 0 ,
	\label{2_30}\\
	\tilde v' + \frac{f \tilde v}{x} &=& \tilde u \left(
	- 1+ \tilde E +
	\frac{\tilde u^2 - \lambda \tilde v^2}{x^2}
	\right) ,
	\label{2_40}\\
	\tilde u' - \frac{f \tilde u}{x} &=& \tilde v \left(
	- 1 - \tilde E +
	\frac{\tilde u^2 - \lambda \tilde v^2}{x^2}
	\right).
	\label{2_50}
\end{eqnarray}
Here, for convenience of making numerical calculations, we have introduced the following dimensionless variables:
$x = r/\lambda_c$,
$
\tilde u=u\sqrt{\Lambda/\lambda_c g},
\tilde v = v\sqrt{\Lambda/\lambda_c g},
\tilde E = \lambda_c E/(\hbar c),
\tilde g^2_{\Lambda} = g \hbar c\lambda_c^2/\Lambda
$, where $\lambda_c= \hbar / (m_f c)$ is the Compton wavelength and $\lambda$ is some combination of the constants
$\alpha, \beta$. The prime denotes differentiation with respect to~$x$.

The total energy density of the monopole under consideration is
\begin{equation}
	\tilde \epsilon =
	\tilde{\epsilon}_m + \tilde \epsilon_s =\frac{1}{\tilde g^2_\Lambda}
	\left[
	\frac{{f'}^2}{ x^2} +
	\frac{\left( f^2 - 1 \right)^2}{2 x^4}
	\right] +
	\left[
	\tilde E \frac{\tilde u^2 + \tilde v^2}{x^2} +
	\frac{\left(\tilde u^2 - \lambda \tilde v^2 \right)^2}{2 x^4}
	\right].
	\label{2_60}
\end{equation}
Here the expressions in the square brackets correspond to the dimensionless energy densities of the non-Abelian gauge fields,
$
\tilde{\epsilon}_m \equiv
\frac{\lambda_c^4 g^2}{\tilde g^2_\Lambda} \epsilon_m
$,
and of the spinor field,
$
\tilde{\epsilon}_s \equiv \frac{\lambda_c^4 g^2}{\tilde g^2_\Lambda} \epsilon_s
$.

Correspondingly, the total energy of the monopole is calculated using the formula
\begin{equation}
	\tilde W_t \equiv \frac{\lambda_c g^2}{\tilde g^2_\Lambda} W_t =
	4 \pi
	\int\limits_0^\infty x^2 \tilde \epsilon d x
	= \left( \tilde{W}_t \right)_m + \left( \tilde{W}_t \right)_{s},
\label{2_70}
\end{equation}
where the energy density $\tilde \epsilon$ is taken from Eq.~\eqref{2_60}, and it is the sum of
the energies of the magnetic field and the nonlinear spinor field.

\section{Numerical results}

Numerical integration of Eqs.~\eqref{2_30}-\eqref{2_50} is carried out using the shooting method with the boundary conditions given for
$x = \delta \ll 1$ (for details see Refs.~\cite{Dzhunushaliev:2020qwf,Dzhunushaliev:2021apa}):
\begin{equation}
	f(\delta) = 1 + \frac{f_2}{2} \delta^2 ,\quad
	f^\prime(\delta) = f_2 \delta , \quad
	\tilde u(\delta) = \tilde u_1 \delta ,\quad
	\tilde v(\delta) = \frac{\tilde v_2}{2} \delta^2 .
\label{3_10}
\end{equation}
The value of the parameter $\tilde v_2 $ can be found from Eqs.~\eqref{2_30}-\eqref{2_50},
$
\tilde v_2 = 2 \tilde u_1 \left(
\tilde E - 1 + \tilde\Lambda \tilde u_1^2
\right)/3
$.
Adjusting the values of the parameters $f_2, \tilde u_1$, one can find the required regular solutions.

Our purpose is to construct a spectrum of solutions for different values of $\lambda$. To do this,
we find regular solutions for different values of the frequency $\tilde E$ with a fixed value of the parameter
$\lambda$. For every value of $\tilde E$, we calculate the total energy  $\tilde W_t$ given by the expression~\eqref{2_70}.
Then we construct the dependence $\tilde W_t(\tilde E)$, using which it is possible to find a mass gap as a minimum of the function $\tilde W_t(\tilde E)$.
Such a procedure is repeated for different values of the parameter  $\lambda$. In the present study we investigate the dependence
of the value of the mass gap $\Delta(\lambda)$ on the parameter $\lambda$. Also, we study the dependence
 $\tilde E_\Delta(\lambda)$ on the same parameter, where the quantity $\tilde E_\Delta$ is defined as the value of the frequency
$\tilde E$ for which the energy spectrum has a minimum, i.e., it is the position of the mass gap on the axis $\tilde E$.

\begin{figure}[H]
\begin{minipage}[t]{.45\linewidth}
\begin{center}
	\includegraphics[width=1\linewidth]{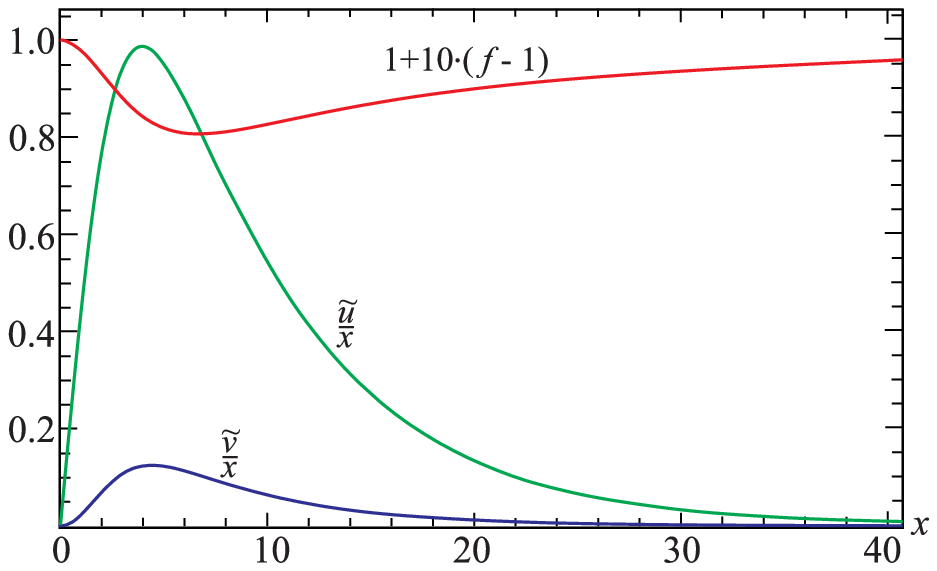}
\end{center}
\vspace{-0.5cm}
\caption{The typical behavior of the functions $f, \tilde u$, and $\tilde v$ of the
``monopole'' solution	for
	$
	\tilde g_{\Lambda} = 0.3354, \lambda = 0.8, \tilde E = 0.99, f_2 = -0.0042694, u_1 = 0.46657
	$.
}
\label{potentials}
\end{minipage} \hfill
\begin{minipage}[t]{.47\linewidth}
\begin{center}
\includegraphics[width=1\linewidth]{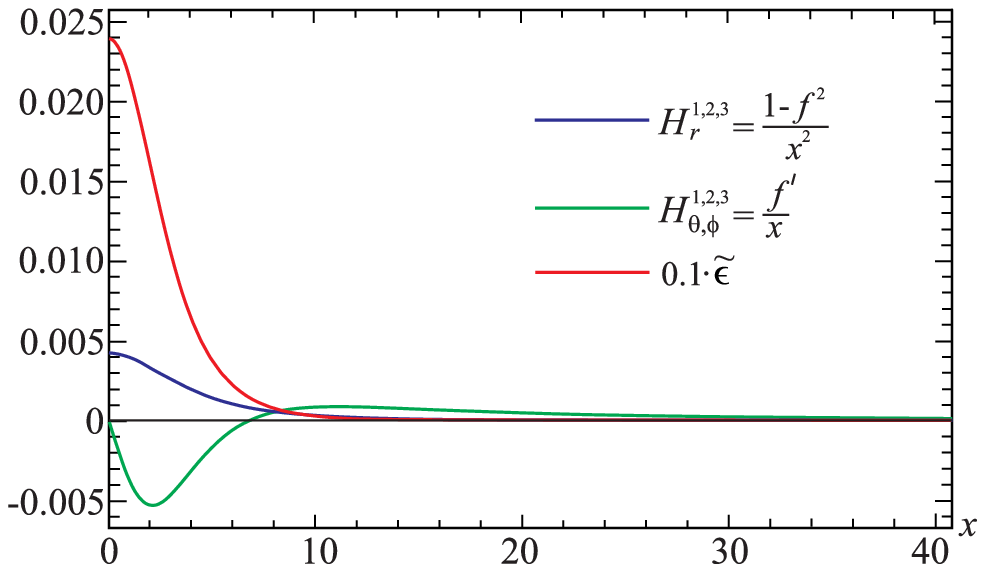}
\end{center}
\vspace{-0.5cm}
\caption{The typical behavior of  the magnetic fields $\tilde H_r, \tilde H_{\varphi, \theta}$ given by Eqs.~\eqref{3_10}-\eqref{3_30}
and the energy density $\tilde \epsilon$ from Eq.~\eqref{2_60} of the ``monopole'' solution for
	$
	\tilde g_{\Lambda} = 0.3354, \lambda = 0.8, \tilde E = 0.99, f_2 = -0.0042694, u_1 = 0.46657
	$.
}
\label{fields}
\end{minipage} \hfill
\end{figure}

The typical solution of the equations~\eqref{2_30}-\eqref{2_50} is given in Fig.~\ref{potentials}, where the profiles of the functions
$f(x), \tilde v(x)$, and $\tilde u(x)$ are given. Fig.~\ref{fields} shows 
the typical distribution of the energy density \eqref{2_60},
as well as the profiles of the  physical components of
the color magnetic fields $H^a_i$  defined as
$
H^a_i=-(1/2)\sqrt{\gamma}\,\epsilon_{i j k} F^{a j k},
$
where $i, j, k$ are space indices and $\gamma$ is the determinant of the spatial metric. This gives the components
\begin{align}
	H^a_r \sim & \frac{1 - f^2}{g r^2},
	\label{3_10}\\
	H^a_{\theta} \sim &\frac{1}{g}f^{\prime},
	\label{3_20}\\
	H^b_{\varphi} \sim &\frac{1}{g}f^{\prime},
	\label{3_30}
\end{align}
where $a=1,2,3$ and we have dropped the dependence on the angular variables.
In these figures,  we use quotation marks for the term ``monopole'' since the asymptotic behavior of the radial magnetic field
 $H^a_r \sim r^{-3}$ differs from that of the  't~Hooft-Polyakov monopole, for which $H^a_r \sim r^{-2}$.

\begin{figure}[H]
	\begin{center}
		\includegraphics[width=.45\linewidth]{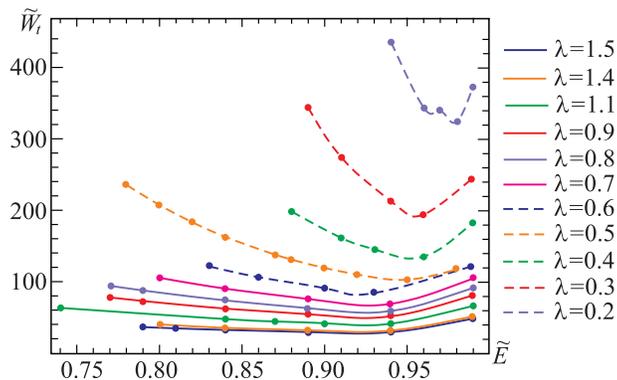}
	\end{center}
	\vspace{-0.5cm}
	\caption{The energy spectra of the ``monopole'' solutions for different values of the nonlinearity parameter $\lambda$.
	}
	\label{spectrums}
\end{figure}

Fig.~\ref{spectrums} shows the dependence of the energy spectrum $\tilde W_t(\tilde E)$ on the parameter $\lambda$.
From this figure, one can conclude that when $\lambda$ increases, the energy spectrum rises, that is the corresponding values of the energy
$\tilde W_t$ increase.

\begin{figure}[H]
\begin{minipage}[t]{.45\linewidth}
\begin{center}
		\includegraphics[width=1\linewidth]{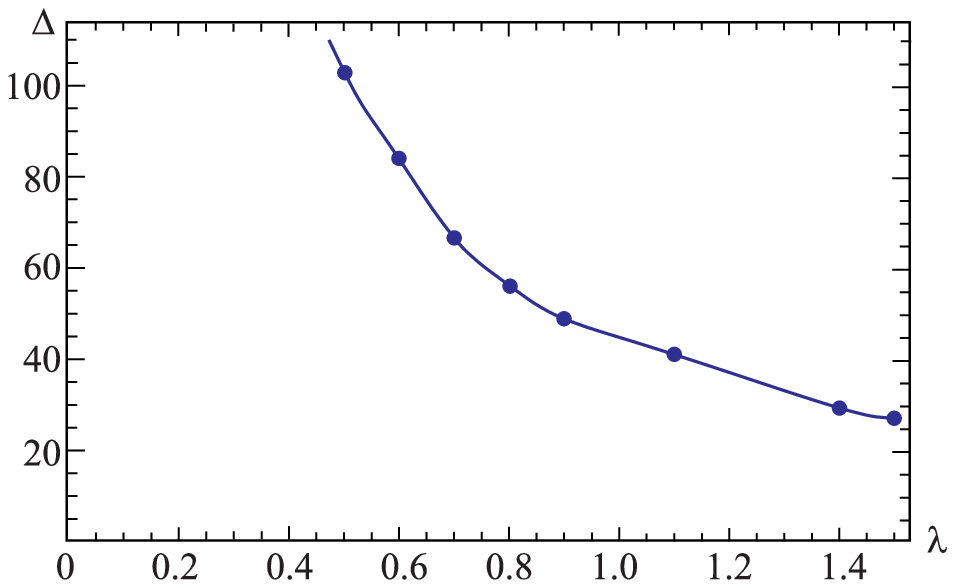}
\end{center}
\vspace{-0.5cm}
\caption{The dependence of the mass gap $\Delta$ on the nonlinearity parameter $\lambda$.
}
\label{Delta_vs_lambda}
\end{minipage} \hfill
\begin{minipage}[t]{.47\linewidth}
\begin{center}
		\includegraphics[width=1\linewidth]{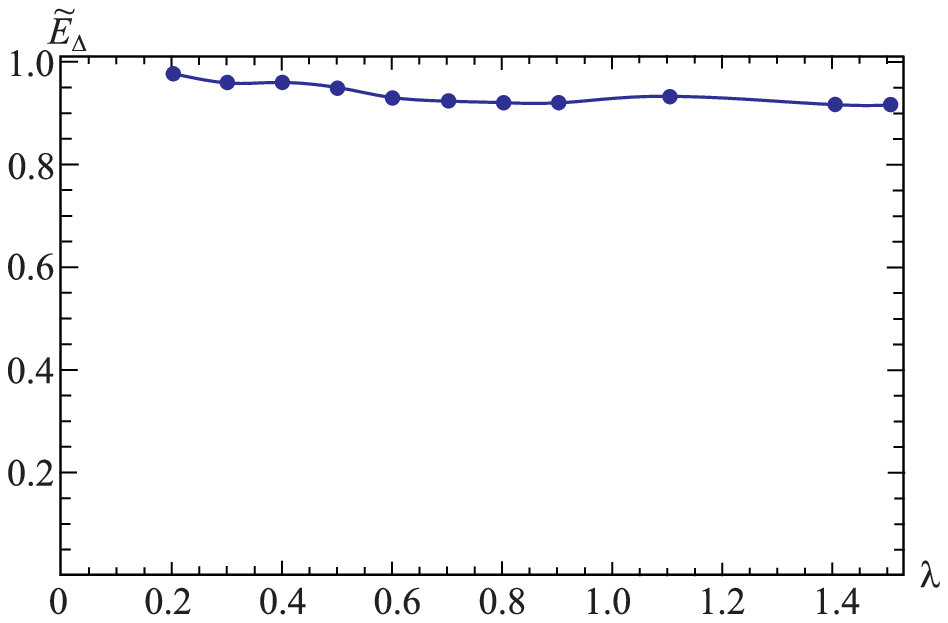}
\end{center}
\vspace{-0.5cm}
\caption{The dependence of the position of the mass gap $\tilde E_\Delta$ on the nonlinearity parameter $\lambda$.
}
\label{position_MG}
	\end{minipage} \hfill
\end{figure}

Our main purpose here is to study the dependence  of the value
of the mass gap $\Delta$ on the nonlinearity parameter  $\lambda$ characterizing the properties of the nonlinear potential of the spinor field $\psi$.
The corresponding results of computations are given in Fig.~\ref{Delta_vs_lambda}.
In turn, Fig.~\ref{position_MG} shows the dependence of the position of the mass gap on the nonlinearity parameter  $\lambda$.
By the term a position of the mass gap, we mean the value of the frequency $\tilde E_\Delta$
for which the energy spectrum has a minimum.
It is of interest to note that the position of the mass gap $\tilde E_\Delta$ does not practically depend on  the nonlinearity parameter
 $\lambda$. One can assume that  in fact this is the case, and deviations from this value are related to errors in numerical calculations.

\section{Conclusions}

In the present paper we have studied the dependence of the value of the mass gap and its position on the parameter describing
the nonlinearity of the spinor field in the Dirac equation. We have found out that the value of the mass gap,
which is assumed to be a global minimum of the energy spectrum of the ``monopole'' solution in SU(2) Yang-Mills theory with a source
in the form of a nonlinear spinor field, depends smoothly on the nonlinearity parameter of the spinor field.
We use the word ``monopole'' in quotation marks because the magnetic field decreases asymptotically as $r^{-3}$,
unlike the  't~Hooft-Polyakov monopole solution where the magnetic field decreases according to the Coulomb law: $r^{-2}$.
It is of interest that the position of the mass gap does not practically depend on the value of the nonlinearity parameter,
at least in the range of values of the spinor field nonlinearity parameter considered here.

Thus, the study indicates that the mass gap in SU(2) Yang-Mills theory with a source
in the form of a nonlinear spinor field does exist for different types of nonlinearity and depends
smoothly on the nonlinearity parameter.

\section*{Acknowledgements}

The work was supported by the program No.~BR10965191 (Complex Research in Nuclear and Radiation Physics, High Energy Physics and Cosmology for the Development of Competitive Technologies)
of the Ministry of Education and Science of the Republic of Kazakhstan. We are also grateful to the Research Group Linkage Programme of the Alexander von Humboldt Foundation for the support of this research.

\end{document}